\documentclass[lettersize,journal]{IEEEtran}
\usepackage{algorithm}
\usepackage{algpseudocode}
\usepackage{array}
\usepackage[caption=false,font=normalsize,labelfont=sf,textfont=sf]{subfig}
\usepackage{textcomp}
\usepackage{stfloats}
\usepackage{url}
\usepackage{verbatim}
\usepackage{graphicx}
\usepackage{subcaption}
\usepackage{subfig}
\usepackage{cite}
\usepackage{amsmath,amssymb,amsfonts}
\usepackage{textcomp}
\usepackage{xcolor}
\usepackage{pgfplots}
\usepackage{pgfplotstable}
\usepackage{filecontents}
\usepackage{tabularx}
\usepackage{tikz}
\usepackage{amsmath, amssymb}
\usepackage{geometry}
\usepackage{bm} 
\usepackage{makecell}
\usepackage{multirow} 
\usepackage{tabularx} 
\usepackage{booktabs}
\usepackage{tikz}
\usetikzlibrary{decorations.pathreplacing,arrows.meta}
\usetikzlibrary{arrows.meta}

\def\BibTeX{{\rm B\kern-.05em{\sc i\kern-.025em b}\kern-.08em
    T\kern-.1667em\lower.7ex\hbox{E}\kern-.125emX}}
\usepackage{balance}

\begin{document}
\title{AoI-driven Anomaly Detection Framework for Communication Networks}
\author{Zia Ul Islam Nasir, Alper Kose, and Berna Ozbek

\thanks{This work was supported by the European Union’s Horizon Europe MSCA-DN programme through the SCION Project under Grant Agreement No. 101072375.

Zia Ul Islam Nasir, Alper Kose, and Berna Ozbek are with the Department of Electrical and Electronics Engineering, Izmir Institute of Technology, Turkiye. (e-mail: zianasir@iyte.edu.tr, alperkose@iyte.edu.tr, bernaozbek@iyte.edu.tr).}}


\maketitle

\begin{abstract}
Anomaly detection in communication networks remains a challenge, particularly in time-sensitive environments where stale information can affect system reliability. In this work, we propose an anomaly detection framework that leverages Age of Information (AoI) as a temporal indicator for identifying abnormal network behaviors. The AoI has been studied for network performance optimization, however, its applicability to  anomaly detection remains unexplored. The proposed framework integrates a feature selection technique based on the estimated Peak Age of Information (ePAoI) with multiple machine learning models for communication networks as a complementary indicator. The evaluation results demonstrate that ePAoI-based feature selection enhances the performance of the anomaly detection framework while significantly reducing the computational complexity.
\end{abstract}


\section{Introduction} 
\IEEEPARstart{O}ver the decades, continuous breakthroughs in communication technologies gave birth to a range of applications with different requirements. In communication networks, machines, sensors, and control entities continuously exchange operational information to support process automation, environmental monitoring, and safety critical applications \cite{perez2024artificial,zhang2022advancements}.  The rapid proliferation of interconnected devices within critical communication infrastructure has introduced significant challenges to ensure reliable network monitoring and timely information delivery. Modern industrial systems, smart cities, and cyber-physical environments increasingly rely on dense sensor networks that continuously collect and transmit operational data for real-time decision making \cite{chataut2023unleashing}. In such time sensitive environments, the detection of anomalies such as sensor faults, node failures, network congestion, and cyber-attacks is essential to maintain system safety, stability, and operational efficiency. However, ensuring timely and reliable information delivery remains challenging due to limited network resources, dynamic traffic conditions, and link disruptions, which can degrade the freshness and usefulness of received data.


The need to maintain fresh information has motivated research across diverse domains, including real-time databases, blockchain systems, and wireless communication networks. In order to quantify information freshness, the Age of Information (AoI) which is defined as elapsed time since the most recently received update was generated at the source \cite{kaul2012real}, is employed. 
Unlike traditional metrics such as latency or throughput, AoI directly captures information freshness, making it particularly relevant for real-time monitoring systems \cite{popovski2024time}. A closely related metric, Peak Age of Information (PAoI), characterizes the maximum age attained immediately before a fresh update is received, offering an additional perspective on worst case information staleness \cite{costa2016age}. Recent studies have extended AoI beyond conventional communication system optimization, highlighting its broader relevance for temporal information quality assessment and its emerging role in privacy aware crowdsensing systems \cite{10870168}. However, despite its growing importance in network performance analysis, its use in anomaly detection remains unexplored.

Meanwhile, anomaly detection in communication networks has evolved from traditional statistical approaches to machine learning (ML) and deep learning-based methods \cite{wang2021machine,huang2025deep,barnard2022robust}. Nevertheless, most existing methods rely primarily on conventional network attributes such as packet loss, delay, and throughput, while often overlooking temporal information freshness as an explicit behavioral feature. This limitation is particularly relevant in dynamic industrial environments, where information staleness itself may reflect abnormal system behavior.

To the best of our knowledge, the integration of AoI-inspired features for anomaly detection in communication networks has not been systematically explored. This paper introduces a novel feature selection approach by incorporating AoI into anomaly detection for communication systems. The proposed framework employs multiple machine learning models based on AoI driven features. It gives a complementary perspective that captures temporal information freshness alongside conventional network centric features, providing additional context for understanding system behavior in dynamic environments.

The main contributions of this paper are summarized as follows:
\begin{itemize}
\item We introduce an AoI-driven framework for anomaly detection in communication networks, complementing conventional network-centric approach.
\item We propose a novel anomaly detection framework by integrating AoI-driven feature selection.
\item The proposed framework is validated using an Industrial Control System dataset.
\end{itemize}

The remainder of this paper is organized as follows. Section II introduces the proposed anomaly detection framework, which is followed by evaluation results and discussion in Section III, and the conclusion is given in Section IV.

\section{Proposed Framework} \label{met}

\begin{figure}[htb]
    \centering
    \includegraphics[width=0.5\textwidth, height=4.9cm]{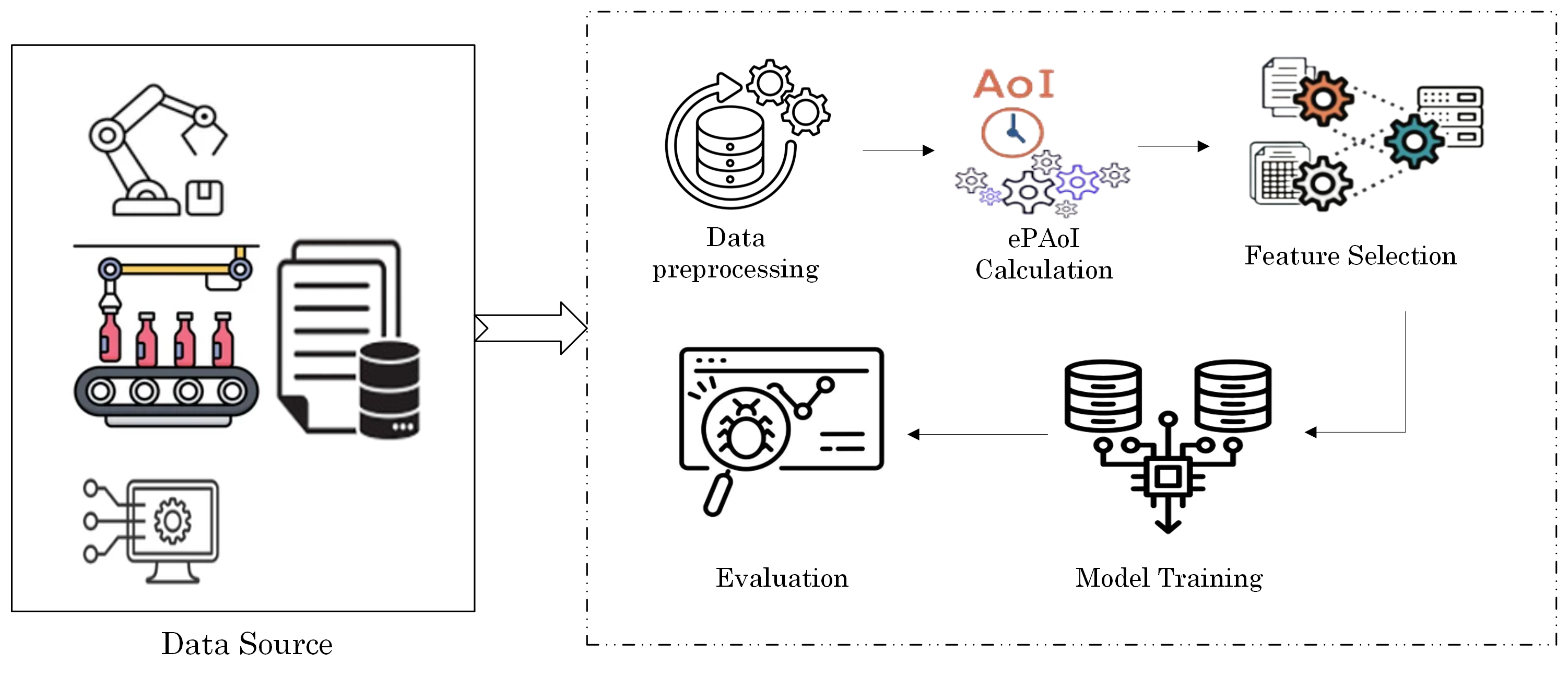}
    \caption{The proposed anomaly detection framework.}
    \label{fig:block}
\end{figure}

This section outlines the process of developing a framework for network anomaly detection using AoI-driven features. The schematic architecture of the proposed framework is illustrated in Fig \ref{fig:block}. The corresponding details including data source and preprocessing, estimated PAoI (ePAoI) calculation, feature selection, and ML based anomaly detection are discussed in the following subsections.

\subsection{Data Source and Preprocessing} 
In this work, we consider an Industrial Control Systems (ICS) \cite{dehlaghi2023anomaly}, in which timely communication between sensing devices and controller is essential for safe and efficient operation. ICS continuously monitors parameters such as temperature, pressure, vibration, and flow rate to regulate physical processes in real time. Transmission delays caused by congestion, failures, or abnormal network behavior may cause controller to act on outdated information, causing degraded performance or safety risks. Therefore, for ICS applications, information freshness becomes as important as transmission reliability.

In the ICS dataset, data streams originate from heterogeneous sources such as sensors, actuators, and controllers. This raw data is preprocessed by handling missing values and normalizing relevant features to ensure data consistency and comparability. The data is then organized by source, destination, and protocol, to enable consistent flow identification and ePAoI computation. Then, we obtain the dataset denoted by \(\mathcal{D} = \{(\bm{x}_n, y_n)\}_{n=1}^N\), where \(N\) is the total number of samples. Each \(\bm{x}_n \in \mathbb{R}^D\) represents the full feature vector with \(D\) is the number of features and \( y_n \in \mathcal{Y}\) is the associated anomaly label where \(\mathcal{Y}= \{0, 1\}\) is the anomaly label set, indicating whether an anomaly has occurred or not.

\subsection{Estimated Peak Age of Information Calculation}
\label{subsec:epaoi}

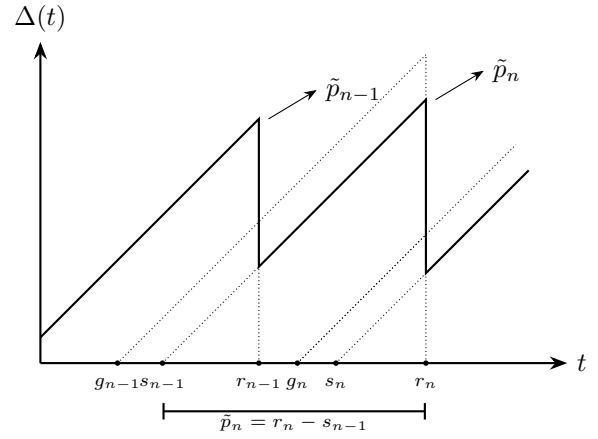
\begin{figure}[!t]
\centering

\begin{tikzpicture}[>=Stealth, scale=0.85]


\def\gA{1.2}
\def\sA{1.9}
\def\rA{3.4}

\def\gB{4.0}
\def\sB{4.6}
\def\rB{6.0}

\def\sP{-0.4}


\draw[->,thick] (0,0) -- (8.2,0) node[right] {$t$};
\draw[->,thick] (0,0) -- (0,5.0) node[above] {$\Delta(t)$};


\draw[thick]
(0,{-\sP})
--
(\rA,{\rA-\sP})
--
(\rA,{\rA-\sA})
--
(\rB,{\rB-\sA})
--
(\rB,{\rB-\sB})
--
(7.6,{7.6-\sB});


\draw[densely dotted]
(\sA,0)
--
(\rA,{\rA-\sA});

\draw[densely dotted]
(\gB,0)
--
(\rB,{\rB-\gB});

\draw[densely dotted]
(\gA,0)
--
(\rB,{\rB-\gA});

\draw[densely dotted]
(\sB,0)
--
(\rB,{\rB-\sB});

\draw[densely dotted]
(\gB,0)
--
(7.4,{7.4-\gB});


\draw[densely dotted]
(\rA,0)--(\rA,{\rA-\sP});

\draw[densely dotted]
(\rB,0)--(\rB,{\rB-\gA});


\foreach \x/\lab in
{
\gA/{g_{n-1}},
\sA/{s_{n-1}},
\rA/{r_{n-1}},
\gB/{g_n},
\sB/{s_n},
\rB/{r_n}
}
{
\fill (\x,0) circle (1.2pt);
\node[below,font=\scriptsize] at (\x,-0.08) {$\lab$};
}


\draw[->]
({\rA+0.15},{\rA-\sP})
--
({\rA+0.90},{\rA-\sP+0.45})
node[right] {${\tilde p}_{n-1}$};

\draw[->]
({\rB+0.15},{\rB-\sA}) 
--
(6.9,4.55)
node[right] {${\tilde p}_n$};


\draw[thick,|-|]
(\sA,-0.75)
--
(\rB,-0.75);

\node[below,font=\scriptsize]
at ({(\sA+\rB)/2},-0.66)
{${\tilde p}_n=r_n-s_{n-1}$};

\end{tikzpicture}

\caption{Illustration of the ePAoI when update generation time is unavailable.}
\label{fig:estimated_paoi}

\end{figure}

Unlike conventional delay oriented metrics, which primarily measure transmission latency, AoI explicitly captures the temporal freshness of the most recently available information at the receiver, thereby providing a richer representation of timeliness. Let $g(t)$ denote the generation time of the freshest update available at the receiver. The AoI at time $t$ can then be expressed as \cite{kaul2012real}:
\begin{equation}
\Delta(t)=t-g(t)
\end{equation}
A closely related metric, PAoI, represents the maximum time elapsed since the generation of the most recently available information before a new update is received. For data stream $n$, the PAoI which represents the maximum value of AoI immediately before the reception of a new update, is expressed as \cite{barakat2019measure, huang2015optimizing}:
\begin{equation}
p_n=r_n-g_{n-1},
\label{eq:true_paoi}
\end{equation}
where $g_{n-1}$ denotes the generation time of the previous status update and $r_n$ denotes the reception time of the current update, as illustrated in Fig.~\ref{fig:estimated_paoi}. Since anomalies in industrial communication networks often manifest through delayed, disrupted, or irregular communication exchanges, PAoI provides a meaningful temporal indicator for anomaly detection, as larger PAoI values naturally indicate prolonged reliance on stale information.

A practical challenge is that the computation of the PAoI requires explicit source side update generation time, which are not available in real dataset. Rather than introducing synthetic assumptions, we estimate it directly from the observed communication chronology. Communication records are first organized into temporally consistent streams defined by the communicating source, destination, and protocol. Since explicit application layer update generation timestamps are unavailable, the earliest observable transmission event of a communication exchange is used as a surrogate reference. The time $s_{n-1}$ refers to the start time or the earliest observable entry of the previous communication record into the communication stream, while $r_n$ denotes the completion of the current communication exchange at the monitoring point. Given the structured and state-driven nature of industrial control communications, this approximation provides a practical estimate of temporal information freshness that can be derived directly from the available network observations. Accordingly, as illustrated in Fig.~\ref{fig:estimated_paoi}, ePAoI is defined as
\begin{equation}
\tilde{p}_n=r_n-s_{n-1}
\label{eq:epaoi}
\end{equation}

After computing the ePAoI, the resulting temporal freshness representation is incorporated into the subsequent feature selection and anomaly detection process.

\subsection{ePAoI-based Feature Selection}

We propose a feature selection scheme centered around the ePAoI, considering it as a cornerstone to identify complementary features that enhance predictive performance. This quantifies the joint contribution of features by capturing their collective discriminative effect rather than their individual relevance. Unlike methods such as mutual information or minimum redundancy-maximum relevance (mRMR), it remains effective when anomalies have weak or heterogeneous label correlations. This approach prioritizes features that jointly interact with the ePAoI to maximize effectiveness of anomaly detection. Then, a subset of features is selected based on the ePAoI, yielding a reduced feature vector \(\bm{\bar x}_n \in \mathbb{R}^K\) with \(K < D\).



For each feature $d$, a feature vector that includes $N$ samples is constructed as \( \bm{x}^{(d)}=[x_1^{(d)},...,x_N^{(d)}] \). Firstly, a discretization process is applied, where features are partitioned into \( B \) bins using equal frequency binning. The number of bins is determined using Sturges's rule:
\begin{equation}
B = \lceil \log_2(N) + 1 \rceil
\label{eq:sturges}
\end{equation}
where \( N \) is the total number of samples. 

The probability that the feature \( \bm{x}^{(d)}\) is into the $b^{\mathrm{th}}$ bin is given by,
\begin{equation}
P \big (X^{(d)}=b\big) = \frac{1}{N} \sum_{n=1}^{N}
\mathcal{I}_{\{x_n^{(d)} = b\}}
\label{eq:marginal_feat_aoi}
\end{equation}
where \( x_n^{(d)} = b \) indicates that the $n^{\mathrm{th}}$ sample of feature $d$ falls in the the $b^{\mathrm{th}}$ bin and $\mathcal{I}_{\{\cdot\}}$ is the indicator function, returning $1$ if the condition is true and $0$ otherwise. 

Analogous to the above, the marginal entropy of the label $y$ is estimated from its empirical class distribution as:
\begin{equation}
H(Y) = - \sum_{c \in \mathcal{Y}} P(Y=c) \log_2 P(Y=c)
\label{eq:label_entropy}
\end{equation}
where $P(Y=c) = \frac{1}{N}\sum_{n=1}^{N} \mathcal{I}_{\{y_n=c\}}$ is the empirical probability
of class $c$.

In order to assess the predictive interaction between the candidate  feature and the ePAoI, we compute the interaction score, which is defined as the joint mutual information between each feature pair and the label $y$:
\begin{align}\label{XX}
S_d &= I(Y; {X}^{(d)}, {X}^{({\tilde p})}) \notag \\
    &= H(Y) - H\big(Y \mid {X}^{(d)}, {X}^{({\tilde p})}\big)
\end{align}
where the conditional entropy \( H(Y \mid {X}^{(d)}, {X}^{(\tilde p)}) \) is given by:
{\small
\begin{equation}
\begin{split}
H\Big(Y \,\big|\, {X}^{(d)}, {X}^{(\tilde p)}\Big)
&= - \sum_{b=1}^{B} \sum_{b'=1}^{B} \sum_{c \in \mathcal{Y}} P(
X^{(d)}=b, X^{(\tilde p)}=b',Y=c) \\
&\quad \times \log_2 \Bigg[
\frac{ P(
X^{(d)}=b, X^{(\tilde p)}=b',Y=c)}
{ P(
X^{(d)}=b, X^{(\tilde p)}=b')} \Bigg]
\end{split}
\label{eq:cond_entropy}
\end{equation}
}
where the joint probabilities are given by
\begin{align}
P(
X^{(d)}=b, X^{(\tilde p)}=b',Y=c) =
 \frac{1}{N} \sum_{n=1}^{N} 
\mathcal{I}_{\{x_n^{(d)}=b,\ x_n^{(\tilde p)}=b',\ y_n=c\}}
\label{eq:joint_label_prob}
\end{align}
and 
\begin{equation}
P \big (X^{(d)}=b, X^{(\tilde p)}=b' \big) = \frac{1}{N} \sum_{n=1}^{N}
\mathcal{I}_{\{x_n^{(d)} = b,\ x_n^{(\tilde p)} = b'\}}
\label{eq:joint_feat_aoi}
\end{equation}

Intuitively, $S_d$ quantifies how much observation of both the feature $\bm{x}^{(d)}$ and the \( \bm{x}^{(\tilde p)} \) reduces uncertainty about the anomaly $y$. It captures feature interactions that are particularly relevant when anomalies exhibit subtle or heterogeneous correlations with labels. A higher score reflects greater predictive relevance, which enables robust and effective feature selection by quantifying the joint contribution of features alongside the ePAoI to anomaly detection.

Finally, the features \( \bm{x}^{(d)}  \quad \forall d \in \{1,\ldots,D\} \) are ranked based on their collaborative score of \( S_d\). The top-ranked features are selected as \( \bm{x}^{(k)}  \quad \forall k \in \{1,\ldots,K\} \) where \( K < D \). Then, for each sample $n$, the reduced feature vector is constructed as \( \bm{\bar x}_n \in \mathbb{R}^K \). This results in a compact yet effective representation that leverages the predictive strength of the ePAoI. The model is then trained to approximate the mapping \(f: \mathcal{X} \rightarrow \mathcal{Y}\), where \(\mathcal{X} = \{\bm{\bar x}_1,...,\bm{\bar x}_N\}\) is the reduced sized feature space.

\begin{algorithm}[t]
\caption{Proposed ePAoI-based Anomaly Detection} 
\label{Algo:Algo1}
\begin{algorithmic}
\State \textbf{Input:} Dataset \(\mathcal{D} = \{(\bm{x}_n, y_n)\}_{n=1}^N\)

\State \textbf{Data Preprocessing}
\State \hspace{1em} Handle missing values 
\State \hspace{1em} Clean and normalize dataset $\mathcal{D}$
\State Calculate ePAoI for each communication stream


\State \textbf{ePAoI-driven Feature Selection}
\For{each feature index $d \in \{1, \dots, D\}$}
    \State \hspace{1em} Compute collaborative score $S_d$ as in (\ref{XX})
\EndFor
\State Rank features by $S_d$ and select top-$K$ features
\State Construct reduced feature vector \( \bm{\bar x}_n \in \mathbb{R}^K \)  for each sample
\State \textbf{Model Training and Evaluation}
\For{each model $M_j$}
    \State \hspace{1em} Train $M_j$ on $\bm{\bar x}_n$ and $\bm{x}^{({\tilde p})}$
    \State \hspace{1em} Evaluate $M_j$ 
\EndFor
\State \Return Trained models $\{M_j\}_{j=1}^5$ and score metrics: accuracy, precision, recall, F1-Score
\end{algorithmic}
\end{algorithm}

\subsection{ML based Anomaly Detection}

A set of ML models \(\{M_j\}_{j=1}^{5}\) including Random Forest (RF), Decision Tree (DT), Artificial Neural Network (ANN), Support Vector Machine (SVM), and Extreme Gradient Boosting (XGBoost) are trained on the reduced feature space, and the performance of each model is evaluated using standard metrics by assigning effective hyper-parameter values and assessing their effects on test data. The step-by-step process is described in Algorithm \ref{Algo:Algo1}.  

From a computational perspective, the proposed approach estimates peak AoI using a lightweight computation for each instance. 
The training complexity of the considered models scales as \(\mathcal{O}(N k \log N)\) for DT, \(\mathcal{O}(T N k \log N)\) for RF with \(T\) trees, \(\mathcal{O}(R N k)\) for XGBoost with \(R\) boosting rounds, \(\mathcal{O}(E N k H)\) for ANN with \(E\) epochs and \(H\) hidden layers, and \(\mathcal{O}(N^3 k)\) for kernel-based SVM. 

\section{PERFORMANCE RESULTS} 

In this section, the performance of the proposed anomaly detection framework is evaluated. The objective is to assess the effectiveness of the models in accurately identifying anomalies. The hyper parameters values of used models are given in Table \ref{tab:hyperparameters}. The proposed anomaly detection approach is implemented using open source Python libraries, including Scikit-learn and TensorFlow.

\begin{table}[h]
\centering
\caption{Hyperparameter Settings for Different Models.}
\label{tab:hyperparameters}
\renewcommand{\arraystretch}{1.2}
\setlength{\tabcolsep}{3pt}
\begin{tabular}{|p{1.6cm}|p{6.4cm}|}
\hline
\textbf{Model} & \textbf{Hyperparameters} \\
\hline
DT & \makecell[l]{Split criterion: entropy \\ min\_samples\_leaf: 1 \\ min\_samples\_split: 2} \\
\hline
RF & \makecell[l]{n\_estimators: 100 \\ Split criterion: gini} \\
\hline
ANN & \makecell[l]{Hidden layers: 2 \\ Activation function: ReLU \\ Optimizer: Adam \\ Learning rate: 0.001 \\ Loss function: binary\_crossentropy \\ Batch size: 32 \\ Epochs: 30} \\
\hline
XGBoost & \makecell[l]{eval\_metric: logloss \\ n\_estimators: 100 \\ max\_depth: 6} \\
\hline
SVM & \makecell[l]{Kernel: polynomial \\ C: 1.0 \\ Gamma: scale} \\
\hline
\end{tabular}
\end{table}

\begin{figure}[htb]
    \centering
\includegraphics[width=0.5\textwidth, height=5cm]{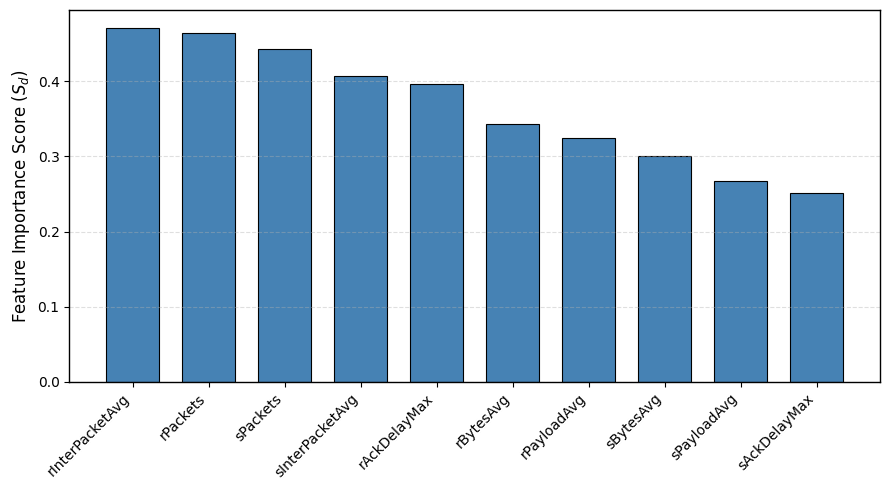} %
    \caption{Ranking of features based on the proposed feature selection method.}
    \label{fig:Feat_rank}
\end{figure}

Initially, the features in dataset are ranked according to their predictive relevance in conjunction with ePAoI as in Eq.(\ref{XX}), which are provided in Fig.~\ref{fig:Feat_rank}. To enhance the predictive performance of the models and reduce the computational complexity, the top K=10 most relevant features identified through the feature selection process, along with the ePAoI feature, are utilized as the input feature set for all considered models. This dimensionality reduction streamlines the training process, decreases computational overhead, and mitigates overfitting risks, thereby improving generalization to unseen data.

Then, the performance of the models is evaluated using standard metrics: accuracy, precision, recall, and F1-score, and the results are given in Table \ref{tab:1}. The results show that the XGBoost achieves the highest performance among all models.

\begin{table}[htb]
\renewcommand{\arraystretch}{1.4}
\centering
\caption{Test results of the models  for the proposed framework.}
\begin{tabular}{llcccc}
\hline
\textbf{Model} & \textbf{Class} & \textbf{Accuracy} & \textbf{Precision} & \textbf{Recall} & \textbf{F1-score} \\
\hline
RF  & Normal     & 0.9479 & 0.9425  & 0.9958 & 0.9684 \\
    & Anomalous  &        & 0.9781  & 0.7544 & 0.8557 \\
\hline
DT  & Normal     & 0.9955 & 0.9967  & 0.9976 & 0.9971 \\
    & Anomalous  &        & 0.9904  & 0.9867 & 0.9885 \\
\hline
ANN & Normal     & 0.9948 & 0.9989  & 0.9947 & 0.9968  \\
    & Anomalous  &        & 0.9790  & 0.9919 &  0.9899 \\
\hline
SVM & Normal     & 0.9741 & 0.9799  & 0.9880 & 0.9839 \\
    & Anomalous  &        & 0.9498  & 0.9180 & 0.9336\\
\hline
XGBoost & Normal     & 0.9975 & 0.9990 & 0.9979 &   0.9985 \\
    & Anomalous  &        & 0.9916 & 0.9960  & 0.9938 \\
\hline
\end{tabular}
\label{tab:1}
\end{table}

In addition to that, a comparative results of the models in terms of True Positive Rate (TPR) and False Positive Rate (FPR) is presented in Table \ref{tab:2}. The results indicate that XGBoost achieves the highest TPR and the lowest FPR among all the models. The high TPR reflects its ability to accurately detect anomalies, while the low FPR minimizes false alarms.

\begin{table}[htb]
\renewcommand{\arraystretch}{1.2}
\centering
\caption{The test result of detection performance for the proposed framework.}
\label{tab:with_without_aoi}
\begin{tabularx}{\linewidth}{Xcc}
\hline
\textbf{Model} & \textbf{TPR } & \textbf{FPR}  \\
\hline
RF     & 0.7544 & 0.0042  \\
DT     & 0.9867 & 0.0024  \\
ANN & 0.9919 & 0.0053  \\
SVM & 0.9180 & 0.0120 \\
XGBoost &\textbf{ 0.9960} & \textbf{0.0021} \\
\hline
\end{tabularx}
\label{tab:2}
\end{table}

\begin{figure}[htb]
    \centering
    \includegraphics[width=0.48\textwidth, height=6cm]{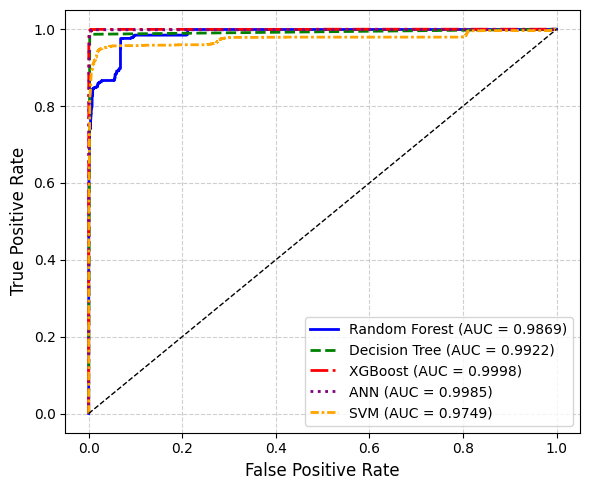} %
    \caption{ROC curves of the models for the proposed framework.}
    \label{fig:roc}
\end{figure}

We also provide the Receiver Operating Characteristic (ROC) curve to evaluate the performance of the proposed framework considering different ML models, as given in Fig. \ref{fig:roc}. As shown in the ROC curve, which is concentrated near the top left corner, the proposed system accurately detects anomalies with very few false positives, highlighting its strong overall performance.

Furthermore, an ablation study is conducted by retraining the models with and without the ePAoI. The corresponding results, presented in Table \ref{tab:3}, demonstrate the effectiveness of incorporating ePAoI into the feature set.  More importantly, it introduces ePAoI to capture information freshness, an important characteristic that has not been considered in existing anomaly detection studies. Beyond improving detection performance, the proposed age aware feature provides additional insight into the freshness of the information used for decision making. 

\begin{table}[htb]
\renewcommand{\arraystretch}{1.2}
\setlength{\tabcolsep}{10pt}
\centering
\caption{Performance Comparison of Models With and Without ePAoI.}
\label{tab:paoi_comparison}
\begin{tabular}{lcccc}
\hline
\multirow{2}{*}{\textbf{Model}} & \multicolumn{2}{c}{\textbf{With ePAoI}} & \multicolumn{2}{c}{\textbf{Without ePAoI}} \\
\cline{2-5}
 & \textbf{F1-Score} & \textbf{FPR} & \textbf{F1-Score} & \textbf{FPR} \\
\hline
RF       & 0.8557 & 0.0042 & 0.8518 & 0.0044 \\
DT       & 0.9885 & 0.0024 & 0.9724 & 0.0054 \\
ANN      & 0.9899 & 0.0027 & 0.9759 & 0.0068 \\
SVM      & 0.9336 & 0.0120 & 0.9250 & 0.0147 \\
XGBoost  & \textbf{0.9938} & \textbf{0.0021} & 0.9860 & 0.0056 \\
\hline
\end{tabular}
\label{tab:3}
\end{table}

We provide the comparison results based on the total inference time and F1-score as in Table \ref{tab:performance_comparison}. Since the benchmark study did not provide inference time, it was  calculated under the same experimental settings to provide a fair evaluation of computational efficiency. Compared with the benchmark study \cite{dehlaghi2023anomaly}, which employs 23 input features, the proposed framework reduces the feature set to only 11 features, lowering the input dimensionality by more than 50\%. Overall, these findings support the use of ePAoI as a complementary age-aware temporal indicator, contributing to computationally efficient anomaly detection while maintaining effective detection performance.

\begin{table}[htbp]
\renewcommand{\arraystretch}{1.2}
\setlength{\tabcolsep}{3pt}
\centering
\renewcommand{\arraystretch}{1.2}
\caption{Performance comparison of the benchmark and the proposed framework in terms of F1-Score and average total inference time.}
\label{tab:performance_comparison}

\begin{tabular}{lcccc}
\hline
\multirow{2}{*}{\textbf{Model}} &
\multicolumn{2}{c}{\textbf{Benchmark \cite{dehlaghi2023anomaly}}} &
\multicolumn{2}{c}{\textbf{Proposed Framework}} \\
\cline{2-5}
& \textbf{F1-Score} & \textbf{Inference Time (ms)}
& \textbf{F1-Score} & \textbf{Inference Time (ms)} \\
\hline
RF       & 0.9885 & 47.2004 & 0.8557 & 38.6163 \\
DT       & 0.9850 & 1.3522  & 0.9885 & 1.1806 \\
ANN      & 0.9860 & 907.1670 & 0.9899 & 846.5540 \\
SVM      & --     & --      & 0.9336 & 1121.0808 \\
XGBoost  & --     & --      & 0.9938 & 28.3133 \\
\hline
\end{tabular}

\end{table}

\section{Conclusion}
In conclusion, this study proposed a pioneering freshness aware framework for network anomaly detection. The integration of AoI into the detection process enables the system to capture both the operational state and the timeliness of the data, thereby enhancing anomaly detection accuracy and improving the model’s generalization. The proposed approach demonstrates high sensitivity to anomalies while remaining computationally efficient and scalable, making it well suited for real-time monitoring in complex, data driven environments. The ablation study also proves the role of ePAoI. In summary, this work provides an intelligent anomaly detection in an ICS environment. Future work will investigate adaptive strategies for AoI computation, along with environment specific parameterization, to further improve robustness across diverse operational settings. In addition, extending the analysis to datasets that capture freshness induced anomalies would enable a more direct and rigorous evaluation of freshness sensitive anomaly detection mechanisms.


\bibliographystyle{IEEEtran}
\bibliography{ref}

\end{document}